\documentclass[11pt,a4paper]{article}
\usepackage[dvips]{graphicx}
\title{
Probing vibrational energy relaxation in proteins
\\
using normal modes
} 
\author{Hiroshi FUJISAKI\footnote{fujisaki@bu.edu}, Lintao BU\footnote{bult@bu.edu}, 
and John E. STRAUB\footnote{straub@bu.edu}
\\
\\
Department of Chemistry, Boston University, 590 Commonwealth Ave.,
\\
Boston, Massachusetts, 02215, USA
}
\begin{document}
\maketitle

\begin{abstract}

Vibrational energy relaxation (VER) of 
a selected mode in cytochrome c (hemeprotein) in vacuum
is studied using two theoretical approaches: One is the equilibrium simulation approach 
with quantum correction factors, and the other is the reduced model approach which 
describes the protein as an ensemble of normal modes coupled with nonlinear coupling elements.
Both methods result in estimates of VER time (sub ps) for a CD stretching 
mode in the protein at room temperature,
that are in accord with the experimental data of Romesberg's group.
The applicability of the two methods is examined through 
a discussion of the validity of Fermi's golden rule on 
which the two methods are based.

\end{abstract}

\tableofcontents 

\section{Introduction}

The harmonic (or normal mode) approximation has been a powerful tool for 
the analysis of few and many-body systems where the 
essential dynamics of the system consists of small oscillations 
about a well-defined mechanically stable structure.
The concept of normal modes (NMs) is appealing in science because 
it provides a simple view for complex systems like solids and proteins.
Though it had been believed that NMs may be too simplistic to analyze the dynamics
of proteins, that is by no means always true: the experimental data 
of neutron scattering for proteins (B factor) indicate that the fluctuations 
for each residue are well represented by a simplified model using NMs \cite{GNN83}.
It was also shown that such a large-amplitude motion as the hinge-bending motion in a protein  
is well described by a NM \cite{MGKW76}.
Importantly, NMs have been used to refine the x-ray structures of proteins \cite{KG90}.
Recently, large proteins or even protein complexes can be analyzed by using NMs \cite{Ma02,LC02,TVFB03}. 

In this chapter, we are concerned with vibrational energy relaxation (VER) in a protein.
This subject is related to our understanding of the functionality of proteins: 
At the most fundamental level, we must understand the energy flow (pathway) 
of an injected energy, that is channeled to do useful work. 
Due to the advance of larser technology, 
time-resolved spectroscopy can detect such energy flow phenomena 
experimentally \cite{MK97}.
To interpret experimental data, and to suggest new experiments,
theoretical approaches and simulations are essential as 
they can provide a detailed view of VER.
However, VER in large molecules itself is still a challenging problem in molecular science \cite{BBS88}.
This is because VER is a typical many-body problem  
and estimations of quantum effects are difficult \cite{Wolynes}.
There is a clear need to test and compare the validity of the existing theoretical methods.
 
We here employ two different methods to estimate the VER rate in a protein, 
cytochrome c (see the next section for details).
One is the classical equilibrium simulation method \cite{SS99} 
with quantum correction factors \cite{BS03,SP01}.
The second is the reduced model approach \cite{FBS04},
which has been recently employed by Leitner's group \cite{Leitner01,Leitner04}.
The latter approach is based on NM concepts, which describes 
VER as energy transfers between NMs mediated by nonlinear resonance \cite{MMK03}. 
We conclude with a discussion of the validity and applicability of such approaches.



\section{Cytochrome c}

\begin{figure}[htbp]
\hfill
\begin{center}
\includegraphics[scale=3.0]{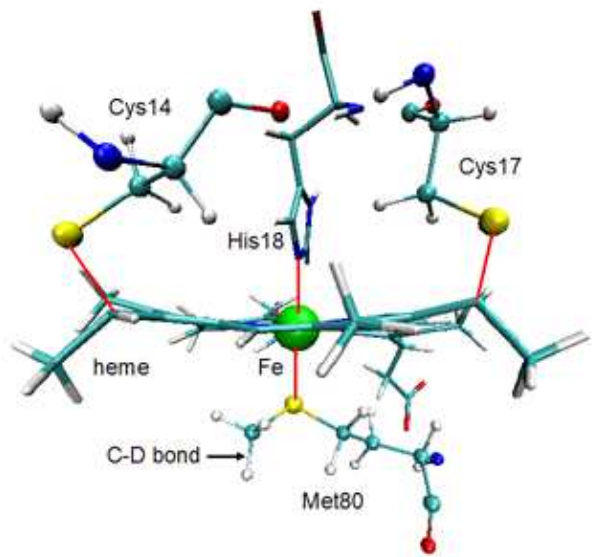}
\end{center}
\caption{
The structure of cytochrome c in the vicinity of the heme group,
showing the thioether linkages and non-planar heme geometry. 
}
\label{fig:cytc}
\end{figure}

Cytochrome c (cyt c) is one of the most thoroughly physicochemically 
characterized metalloproteins \cite{cytc,simulation}. It consists of a single polypeptide 
chain  of 104 amino acid residues and is organized into a series of five 
$\alpha$-helices and six $\beta$-turns. The heme active site in cyt c consists of 
a 6-coordinate low-spin iron that binds His18 and Met80 as the axial ligands. 
In addition, two cysteines (Cys14 and Cys17) are covalently bonded through thioether 
bridges to the heme (see Fig.~\ref{fig:cytc}). Crystal structures of cyt c show that the heme group, 
which is located in a groove and almost completely buried inside the protein, 
is non-planar and somewhat distorted into a saddle-shape geometry. The reduced protein, 
ferrocytochrome c (ferrocyt c), is relatively compact and very stable, due to the fact 
that the heme group is neutral.

The vibrational mode we have chosen for study is the isotopically labeled CD stretch in 
the terminal methyl group of the residue Met80, which is covalently bonded to Fe in heme
(see Fig.~\ref{fig:cytc}). Our simulation model approximates the protein synthesized 
by Romesberg's group \cite{CJR01}
though their protein contains three deuteriums in Met80 (Met80-3D). 
The CH and CD stretching bands are located near 3000 cm$^{-1}$ 
and 2200 cm$^{-1}$, respectively. In contrast with the modeling of photolyzed CO in myoglobin 
\cite{SS99}, 
essentially a diatomic molecule in a protein ``solvent,'' we are interested in the relaxation 
of a selected vibrational mode of a protein. As a result, the modeling is more challenging: 
There is no clean separation between the system and bath modes because 
the CD bond is strongly connected to the environment. 

\section{Quantum correction factor approach}

The classical Landau-Teller-Zwanzig theory of VER is attractive in that it allows us to base our estimate of the VER rate on a classical force autocorrelation function that contains the interaction coupling between the system and bath modes to all orders.  The Hamiltonian for such a system is of the Caldeira-Leggett-Zwanzig form, where ``bath'' coordinates are represented as normal modes of the {\it bath alone.}\footnote{
As shown below, the {\it classical} LTZ formula can be considered as a classical limit 
of the quantum mechanical population relaxation rate $1/T_1$. 
This result is derived by using {\it both} Fermi's golden rule and 
Bader-Berne theory \cite{BB94}. Though the transition rate $k_{n \rightarrow n-1}$ itself 
can be derived without any assumption on the bath Hamiltonian, the 
Bader-Berne result stems from the assumption that the bath Hamiltonian is 
an ensemble of harmonic oscillators.} 
 The relaxing oscillator is introduced as a local ``system'' mode, coupled to the bath at all orders, including ``bi-linear'' coupling.  

Efforts have been made to introduce quantum effects through the use of ``quantum correction factors.''  The dynamics of the classical system are computed, and the quantum effects are added {\it a posteriori} in a manner that accounts for the equilibrium quantum statistical distribution of the contributing quantum mechanical degrees of freedom.   This approach is summarized below and applied to estimate the rate of VER for the CD bond in the terminal methyl group of Met 80 in cytochrome c.

\subsection{Fermi's golden rule}

Our starting point for computing the rate of vibrational energy relaxation of the CD stretching mode in cytochrome c is Fermi's ``golden rule'' formula.  The vibrational population relaxation rate can be written as \cite{FBS04,BB94}
\begin{eqnarray}
\frac{1}{T_1}
=
 \frac{\tanh({\beta \hbar \omega_S/2})}{\beta \hbar \omega_S/2}  
\int_0^{\infty} dt \, \cos (\omega_S t)  \, \zeta_{\rm qm}(t)
=
 \frac{\tanh({\beta \hbar \omega_S/2})}{\beta \hbar \omega_S/2}  
\tilde{\zeta}_{\rm qm}(\omega_S)
\label{eq:Fermi}
\end{eqnarray}
where the force-force correlation function $\zeta_{\rm qm}(t)$ is defined as 
\begin{equation}
\zeta_{\rm qm}(t)=
\frac{\beta}{2 m_S} \langle {\cal F}(t) {\cal F}(0) + {\cal F}(0) {\cal F}(t) \rangle_{\rm qm},
\end{equation}
its Fourier transform is $\tilde{\zeta}_{\rm qm}(\omega)$,
${\cal F}(t)$ is the quantum mechanical force applied to the system mode considered,
$m_S$ is the system (reduced) mass, 
$\omega_S$ is the system frequency,
$\beta$ is an inverse temperature, 
and the above bracket means a quantum mechanical average.
Note that in the classical limit $\hbar \rightarrow 0$, the prefactor in 
front of the integral in Eq.~(\ref{eq:Fermi}) becomes unity, and 
the expression reduces to the well-known classical VER formula.
The issue is that this limit does not represent well the 
VER for high frequency modes because of quantum effects (fluctuation),
whereas it is difficult to calculate $\zeta_{\rm qm}(t)$.

Rather than using the population relaxation rate $1/T_1$, we could compute the rate of transition between pairs of vibrational quantum states 
\begin{eqnarray}
  k_{n \rightarrow n-1}^{\rm qm} = \frac{2n}{\beta \hbar \omega_S [1 + e^{-\beta \hbar \omega_S}]}
\tilde{\zeta}_{\rm qm}(\omega)
\end{eqnarray}
where $n$ is the vibrational quantum number.  In the limit that $\beta \hbar \omega_S \gg 1$ as we consider here, the splitting between vibrational levels is large compared with the thermal energy.  At equilibrium, the system oscillator will be ground state dominated, and we find that 
\begin{eqnarray}
\frac{1}{T_1}
\simeq 
\frac{2\tilde\zeta_{\rm qm}(\omega_S)}{\beta \hbar \omega_S} \simeq k_{1 \rightarrow 0}^{\rm qm} .
\end{eqnarray}
For such a system, we are free to consider the rate of vibrational relaxation in terms of the ensemble averaged relaxation rate $1/T_1$ or the microscopic rate constant $k_{1 \rightarrow 0}^{\rm qm}$ --- the results will be equivalent.

In the limit that $\beta \hbar \omega_S \rightarrow 0$, on the other hand, 
the splitting between states becomes much smaller than the thermal energy and the results are {\it not} equivalent.  The rate constant $k_{1 \rightarrow 0}^{qm}$ diverges, while the population relaxation rate $1/T_1$ is well behaved
\begin{eqnarray}
\frac{1}{T_1}
\simeq 
\tilde\zeta_{\rm qm}(\omega_S).
\end{eqnarray}
In this work, we will present our results in terms of $1/T_1$.

\subsection{Quantum correction factor}

While $\zeta_{\rm qm}(t)$ is difficult to compute for all but the simplest systems, it is often possible to compute the classical analog 
\begin{equation}
\zeta_{\rm cl}(t)=
\frac{\beta}{m_S} \langle {\cal F}(t) {\cal F}(0) \rangle_{\rm cl}
\end{equation}
for highly non-linear systems consisting of thousands of atoms.  The above bracket denotes a classical ensemble average.  The challenge is to relate the quantum mechanical correlation function to its classical analog.  An approach explored by Skinner and coworkers has proved to be quite productive \cite{SP01}.  It involves relating the spectral density of the quantum system to that of the analogous classical system as 
\begin{equation}
\tilde\zeta_{\rm qm}(\omega_S) = Q(\omega_S) \tilde\zeta_{\rm cl}(\omega_S) 
\end{equation}
where $Q(\omega_S)$ is referred to as the ``quantum correction factor (QCF).''  The QCF must obey detailed balance $Q(\omega) = Q(-\omega) e^{\beta \hbar \omega}$ and satisfy the ``classical'' limit that as $\beta \hbar \omega$ becomes small, the QCF approaches unity.  Using this result, we may rewrite Eq.~(\ref{eq:Fermi}) as
\begin{eqnarray}
\frac{1}{T_1^{\rm QCF}}
\simeq 
\frac{Q(\omega_S)}{\beta \hbar \omega_S} 
\tilde{\zeta}_{\rm cl}(\omega_S)
\label{eq:qcf}
\end{eqnarray}
Note that the classical VER rate is defined 
as $1/T_1^{\rm cl} \equiv \tilde{\zeta}_{\rm cl}(\omega_S)$.

\begin{figure}[htbp]
\hfill
\begin{center}
\includegraphics[scale=1.5]{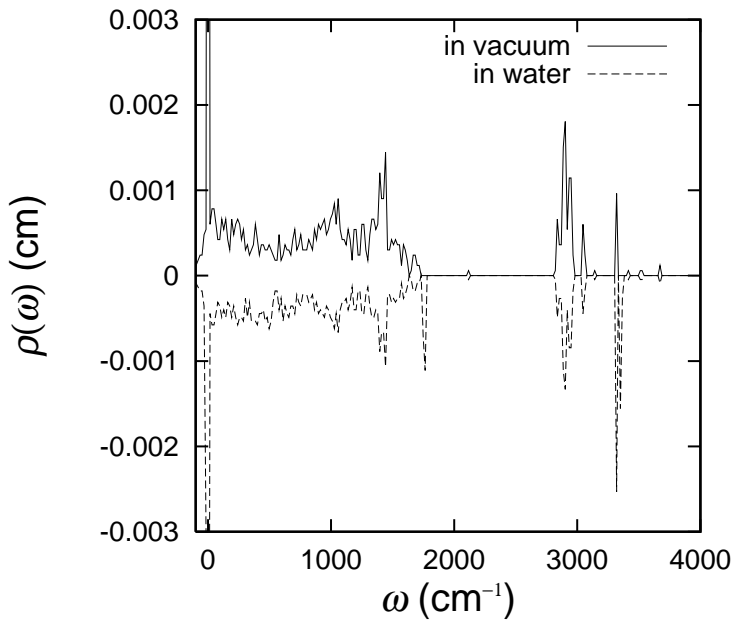}
\end{center}
\caption{
Density of states $\rho(\omega)$ for cytochrome c in vacuum (solid line) and in water (dashed line) at 300K calculated 
by INM analysis.
}
\label{fig:dos}
\end{figure}

The QCF for a one phonon relaxation mechanism is
\begin{equation}
Q_H(\omega)
=
\frac{\beta \hbar \omega}{1-e^{-\beta \hbar \omega}}.
\end{equation}
However, as the CD streching mode falls in the transparent region of the DOS (Fig.~\ref{fig:dos}),
a 1:1 Fermi resonance (linear resonance) is not the possible mechanism of VER.
As such, the lowest order mechanism available for the VER of the CD mode
should involve two phonons.

We have employed Skinner's QCF approach for two-phonon relaxation \cite{SP01}.  
If the assumed two-phonon mechanism assumes that two lower frequency bath modes, having frequencies $\omega_A$ and $\omega_S - \omega_A$, are each excited by one quantum of vibrational energy, the appropriate QCF is 
\begin{equation}
Q_{HH}(\omega_S)
= Q_H(\omega_A) Q_H(\omega_S-\omega_A).
\label{eq:QCFHH}
\end{equation}
Alternatively, if the assumed two phonon mechanism is one that leads to the excitation of one bath vibrational mode of frequency $\omega_A$, with the remaining energy $\hbar (\omega_S - \omega_A)$ being transferred to lower frequency bath rotational and translational modes, the appropriate QCF is
\begin{equation}
Q_{H-HS}(\omega_S)
=
Q_H(\omega_A) \sqrt{Q_{H}(\omega_S-\omega_A)} e^{\beta \hbar (\omega_S-\omega_A)/4},
\label{eq:QCFHHS}
\end{equation}
The functions $Q_H, Q_{HH}, Q_{H-HS}$ 
are called the harmonic, harmonic-harmonic, harmonic-harmonic-Schofield QCF,
respectively.  

\subsection{Normal mode calculations for cytochrome c}

To compute the QCF requires a knowledge of, or guess at, the mechanism of vibrational energy relaxation.  Likely bath modes must be identified and a combination of those modes must meet a resonance condition enforced by the conservation of energy in the transition.    The candidate modes are identified using quenched normal mode (QNM) or instantaneous normal mode (INM) calculations.

In Fig.~\ref{fig:dos}, we show the density of states (DOS) for cyt c in vacuum 
and in water at 300K. These are the INM spectra,
so they contain some negative (actually imaginary) components.
The basic structure of this DOS is similar to that of other proteins like myoglobin \cite{SS99,MMK03}.
The librational and torsional motions are embedded in lower frequency regions below 
2000 cm$^{-1}$, and vibrational motions are located in higher frequency regions 
around 3000 cm$^{-1}$. There is a transparent region between 
2000 cm$^{-1}$ and 3000 cm$^{-1}$; the peak 
due to the CD mode falls in this region near 2200 cm$^{-1}$. 
The VER of this CD mode is our target in this study.
Note, furthermore, that the spectra in vacuum and in water are very similar:
This indicates that water solvent might not affect the simulation results. 
This conjecture will be confirmed later.

\subsection{Application to VER of the CD bond in cytochrome c}

Bu and Straub \cite{BS03} employed the QCF approach to estimate the VER rate of a CD bond 
in the terminal methyl group of Met80 in cyt c (Fig.~\ref{fig:cytc}). 
Their calculations were done using the program CHARMM \cite{CHARMM},
and cyt c was surrounded by water molecules at 300K.
Compared to this,
we have used molecular dynamics simulations of cyt c {\it in vacuum} at 300K 
to compute the classical autocorrelation function for the force acting on the same CD bond.
The results have been used to make estimates of both $1/T_1^{\rm cl}$ and $1/T_1^{\rm QCF}$.

\begin{figure}[htbp]
\hfill
\begin{center}
\begin{minipage}{.32\linewidth}
\includegraphics[scale=0.7]{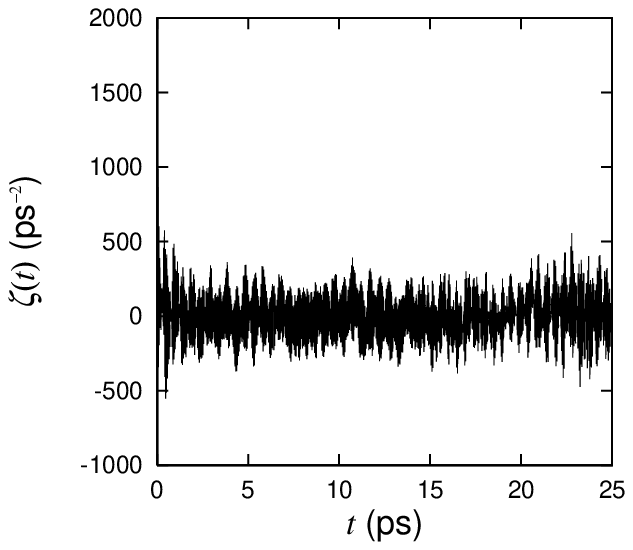}
\end{minipage}
\begin{minipage}{.32\linewidth}
\includegraphics[scale=0.7]{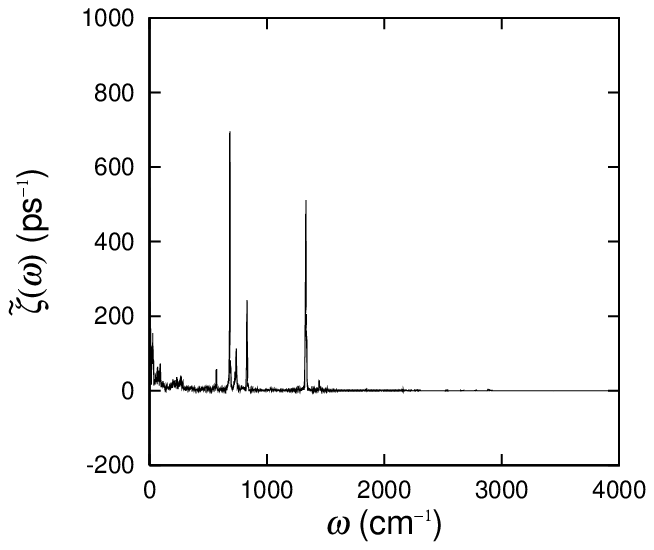}
\end{minipage}
\begin{minipage}{.32\linewidth}
\includegraphics[scale=0.7]{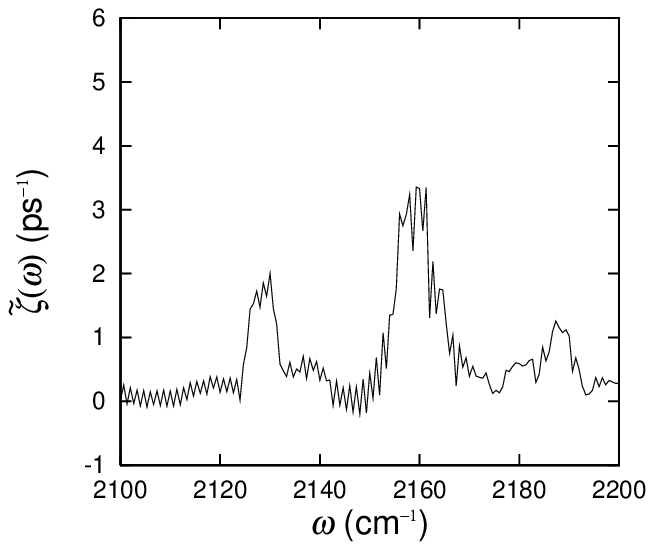}
\end{minipage}
\begin{minipage}{.32\linewidth}
\includegraphics[scale=0.7]{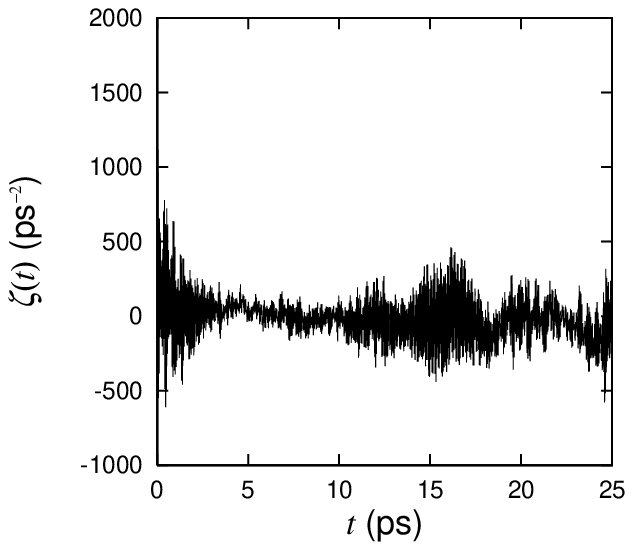}
\end{minipage}
\begin{minipage}{.32\linewidth}
\includegraphics[scale=0.7]{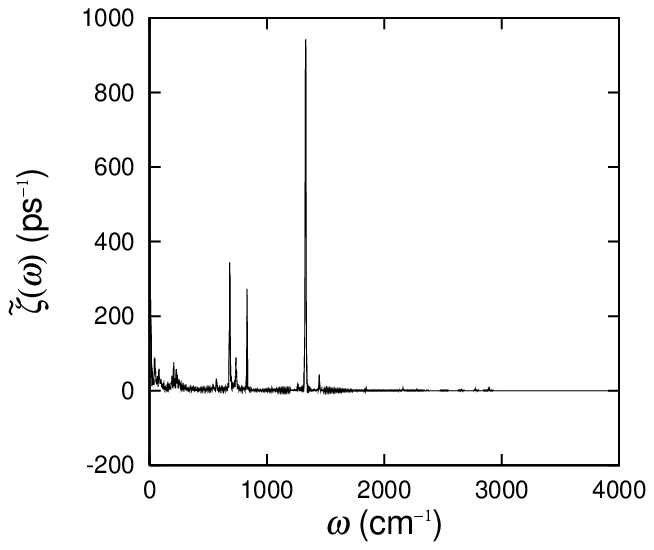}
\end{minipage}
\begin{minipage}{.32\linewidth}
\includegraphics[scale=0.7]{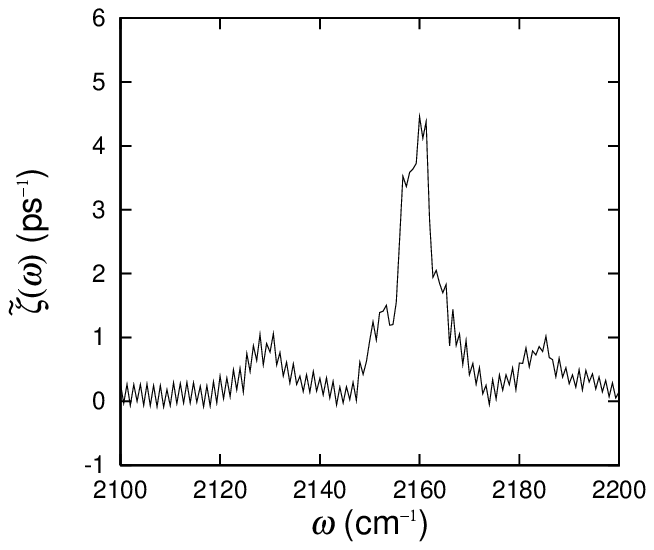}
\end{minipage}
\begin{minipage}{.32\linewidth}
\includegraphics[scale=0.7]{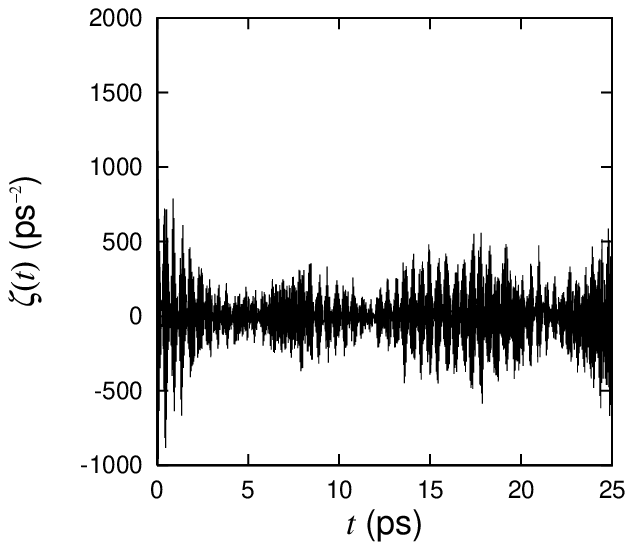}
\end{minipage}
\begin{minipage}{.32\linewidth}
\includegraphics[scale=0.7]{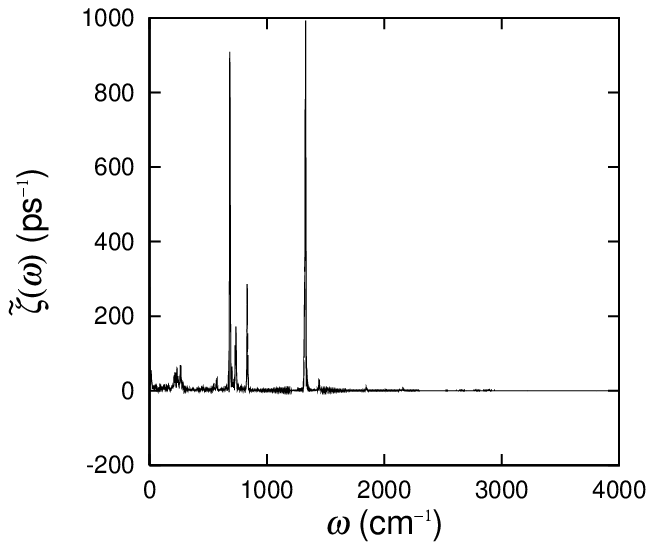}
\end{minipage}
\begin{minipage}{.32\linewidth}
\includegraphics[scale=0.7]{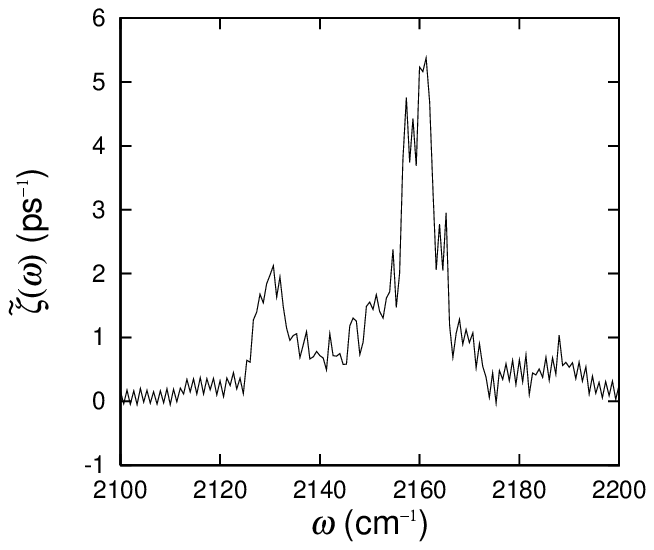}
\end{minipage}
\begin{minipage}{.32\linewidth}
\includegraphics[scale=0.7]{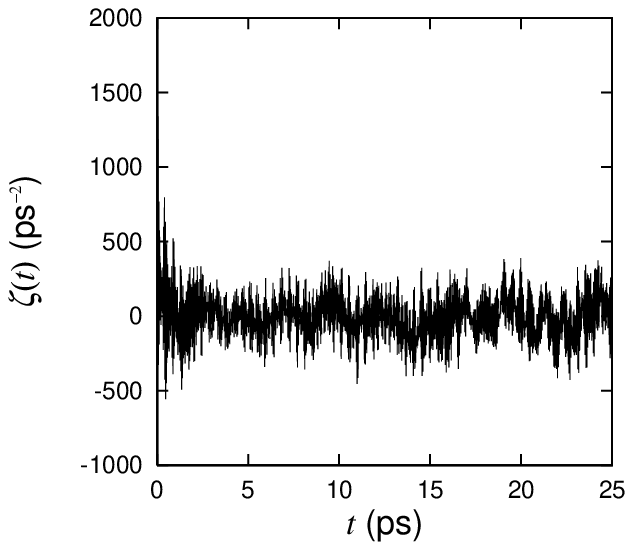}
\end{minipage}
\begin{minipage}{.32\linewidth}
\includegraphics[scale=0.7]{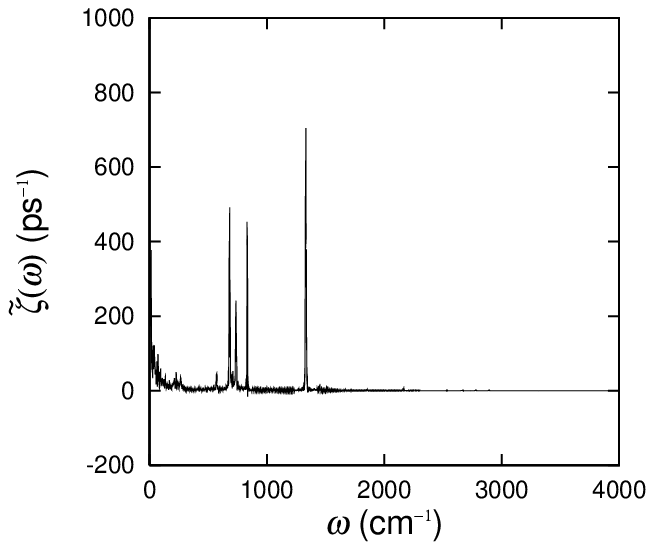}
\end{minipage}
\begin{minipage}{.32\linewidth}
\includegraphics[scale=0.7]{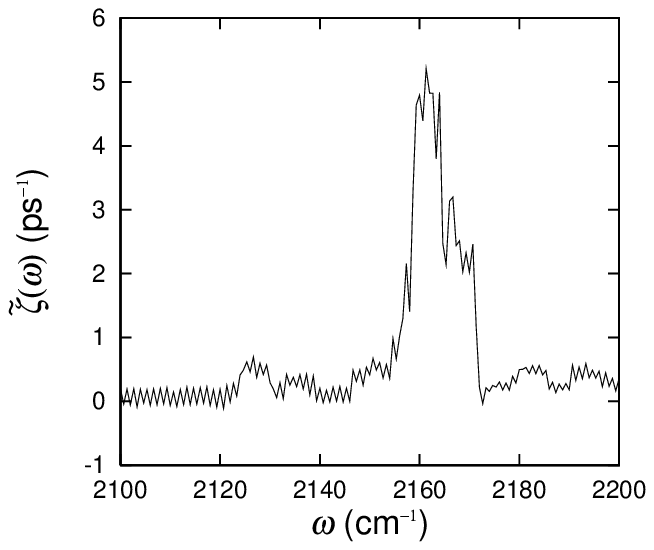}
\end{minipage}
\end{center}
\caption{
Left: Classical data for the force-force correlation function.
Middle: Fourier spectra for the correlation function. 
Right: Magnification of the middle figures around the 
CD bond frequency.
These data are taken from four different trajectories of the 
equilibrium simulation.
}
\label{fig:classical}
\end{figure}

In Fig.~\ref{fig:classical}, the force autocorrelation function 
and its power spectrum are shown for four different trajectories.
We have observed that the force fluctuation and the magnitude of the 
power spectrum for cyt c in vacuum is very similar to 
those computed for cyt c in water.
We conclude that the effects of water on the VER rate 
are negligible.  With the CD bond frequency $\omega_S=2133$ cm$^{-1}$, 
we find $1/T_1^{\rm cl}=\tilde{\zeta}_{\rm cl}(\omega_S) \simeq $ 1 ps$^{-1}$, 
i.e. the classical VER time is about 1 ps.

To apply QCFs for two-phonon relaxation, 
Eqs.~(\ref{eq:QCFHH}) and (\ref{eq:QCFHHS}), 
to this situation, 
we need to know $\omega_A$.
We have found that the CD mode is strongly resonant with two lower frequency modes, 
1655th (685.48 cm$^{-1}$) and 3823rd (1443.54 cm$^{-1}$) modes because 
$|\omega_S-\omega_{1655}-\omega_{3823}| =0.03$ cm$^{-1}$.
We might be able to choose $\omega_A=1443.54$ cm$^{-1}$ or 685.48 cm$^{-1}$.

\begin{figure}[htbp]
\hfill
\begin{center}
\begin{minipage}{.42\linewidth}
\includegraphics[scale=1.1]{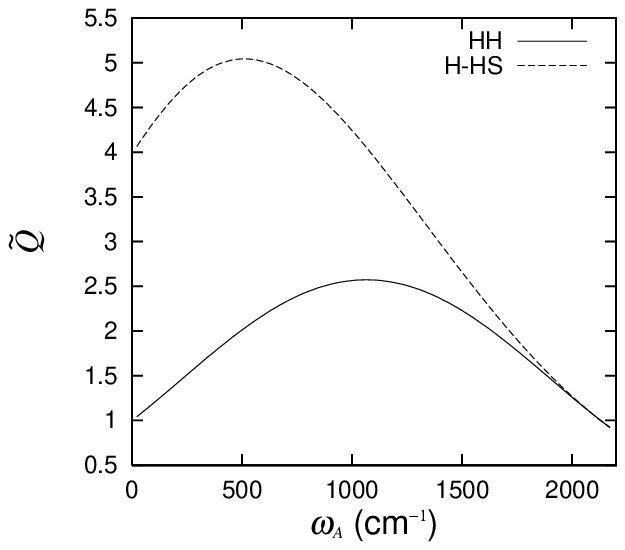}
\end{minipage}
\hspace{1cm}
\begin{minipage}{.42\linewidth}
\includegraphics[scale=1.1]{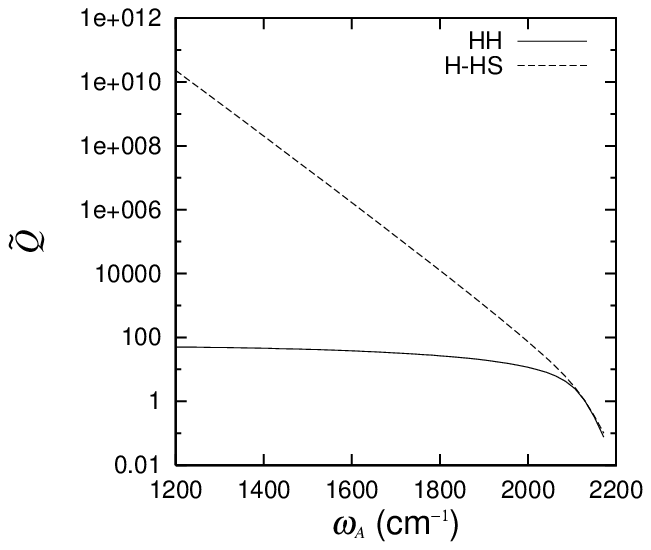}
\end{minipage}
\end{center}
\caption{
Normalized harmonic-harmonic (HH) and harmonic-harmonic-Schofield (H-HS) QCF 
at 300K (left) and at 15K (right).
}
\label{fig:QCF}
\end{figure}

In Fig.~\ref{fig:QCF}, we show $\omega_A$ dependence of the 
normalized QCF, i.e. $\tilde{Q}=Q/(\beta \hbar \omega_S)=T_1^{\rm cl}/T_1^{\rm QCF}$ at 300K 
and at 15K.
If we choose $\omega_A=1443.54$ cm$^{-1}$ at 300K, 
$\tilde{Q}=2.3$ for the harmonic-harmonic QCF and 
2.8 for the harmonic-harmonic-Schofield QCF.
Thus we have $T_1^{\rm QCF}=T_1^{\rm cl}/\tilde{Q}=0.3 \sim 0.4$ ps.
It is interesting to note $\tilde{Q}$ at 15K varies significantly 
depending on the QCF employed.
We will discuss this feature later.

\subsection{Fluctuation of the CD bond frequency}
\label{sec:freq-fluc}

\begin{figure}[htbp]
\hfill
\begin{center}
\includegraphics[scale=1.5]{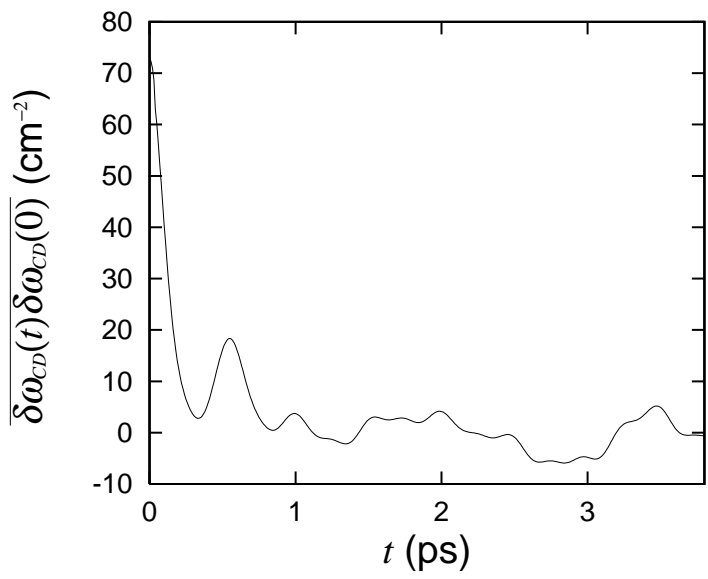}
\end{center}
\caption{
The frequency autocorrelation function. The data up to 20 ps 
were used to calculate the correlation function.
}
\label{fig:freq-fluc}
\end{figure}

We have discussed the fluctuation of the frequency for the CD bond \cite{BS03}.
In the equilibrium simulation, the 
instantaneous normal mode analysis has been employed for each instant of time
to generate a time series $\omega_{CD}(t)$ for the CD bond frequency. 
From this time series, we can calculate
the frequency autocorrelation function
\begin{equation}
C(t)=\overline{\delta \omega_{CD}(t) \delta \omega_{CD}(0)}
=\frac{1}{T}\int_0^T d\tau \delta \omega_{CD}(t+\tau) \delta \omega_{CD}(\tau)
\label{eq:fre-auto}
\end{equation}
where the overline means a long time ($T$) average, and 
$\delta \omega_{CD}(t) = \omega_{CD}(t) -\overline{\omega_{CD}(t)}$.
The correlation time is 
defined as 
\begin{equation}
\tau_c = \frac{1}{(\Delta \omega)^2} \int_0^{\infty} C(t) dt
\end{equation}
where $(\Delta \omega)^2=C(0)$. 
From Fig.~\ref{fig:freq-fluc}, we found 
$\Delta \omega \simeq 8.5$ cm$^{-1}$ and $\tau_c \simeq 0.2$ ps.
Since $\Delta \omega \tau_c \ll 1$, 
according to Kubo's analysis \cite{Kubo}, 
the lineshape should be homogeneously broadened, i.e., 
its shape is Lorentzian. 
This is also the case for cyt c in water \cite{BS03}.
We also confirmed that 
the potential barrier of the methyl group to rotate 
is significantly greater than the thermal energy (barrier height $\simeq$ 3 kcal/mol $>$ 
thermal energy $\simeq$ 0.6 kcal/mol) 
so that we do not expect inhomogeneity in the line shape. 

These results support the validity of employing a normal mode type study 
of VER in cyt c 
as the structure of cyt c is rather rigid around the CD bond and 
the dynamics of the bond, on the scale of VER, should be well modeled by NMs.
Of course, to describe VER among NMs, we must include 
nonlinear coupling terms. 
In the next section, 
we will discuss this reduced model approach for VER in cyt c.

\section{Reduced model approach}

The QCF approach is attractive in that it allows us to base our estimate of the VER rate on a force autocorrelation function that contains the non-linearity of the coupling between the system and bath modes to all orders.  To obtain an estimate of the VER based on a more accurate representation of the system's quantum dynamics, we expand the potential as a Taylor series in the normal modes of the of the system and bath.  In this representation, the system and bath coordinates are ``normal modes;''  to second order in the  expansion of the potential energy, the system and bath modes are uncoupled and non-interacting.  The interaction ``coupling'' between the system and bath modes first appears at third-order.   In this section, we describe a perturbation theory estimate of the rate of VER of the CD mode that represents the system-bath coupling to lowest, third-order  in the system and bath coordinates.

\subsection{Reduced model for a protein}

The reduced model approach utilizes the normal mode picture of a protein,
expanding the residual term perturbatively as \cite{SO98}
\begin{eqnarray}
{\cal H} &=& 
{\cal H}_S+{\cal H}_B+{\cal V}_3 +{\cal V}_4 + \cdots,
\\
{\cal H}_S 
&=& \frac{p_S^2}{2}+\frac{\omega_S^2}{2}q_S^2,
\\
{\cal H}_B 
&=& \sum_k \frac{p_k^2}{2}+\frac{\omega_k^2}{2}q_k^2,
\\
{\cal V}_3 
&=&
\sum_{k,l,m} G_{klm} q_k q_l q_m,
\\
{\cal V}_4 
&=&
\sum_{k,l,m,n} H_{klmn} q_k q_l q_m q_n.
\label{eq:4th}
\end{eqnarray}
Thus the force applied to the system mode is 
\begin{equation}
{\cal F} =-\frac{\partial {\cal V}}{\partial q_S}
=- 3 \sum_{k,l} G_{S,k,l} q_k q_l -4 \sum_{k,l,m} H_{S,k,l,m} q_k q_l q_m + \cdots
\end{equation}
where we have used the permutation symmetry of $G_{klm}$ and $H_{klmn}$.
If it is enough to include the lowest order terms proportional to 
$G_{klm}$,
substituting them into Fermi's golden rule Eq.~(\ref{eq:Fermi}),
we can derive an approximate VER rate as \cite{FBS04}
\begin{eqnarray}
\frac{1}{T_1}
\simeq 
\frac{1}{m_S \hbar \omega_S}
\frac{1-e^{-\beta \hbar \omega_S}}{1+e^{-\beta \hbar \omega_S}} 
\sum_{k,l} 
\left[
\frac{\gamma \zeta^{(+)}_{k,l}}{\gamma^2+(\omega_k+\omega_{l}-\omega_S)^2}
+
\frac{\gamma \zeta^{(+)}_{k,l}}{\gamma^2+(\omega_k+\omega_{l}+\omega_S)^2}
\nonumber
\right.
\\
\left.
+
\frac{\gamma \zeta^{(-)}_{k,l}}{\gamma^2+(\omega_k-\omega_{l}-\omega_S)^2}
+
\frac{\gamma \zeta^{(-)}_{k,l}}{\gamma^2+(\omega_k-\omega_{l}+\omega_S)^2}
\right]
\label{eq:rate}
\end{eqnarray}
where we have included a width parameter $\gamma$ to broaden 
a delta function, and defined the following   
\begin{eqnarray}
\zeta_{k,l}^{(+)}
&=&
\frac{\hbar^2}{2}
\frac{(A_{k,l}^{(2)})^2}{\omega_k \omega_{l}}
(1+n_k +n_{l} + 2 n_k n_{l}),
\\
\zeta_{k,l}^{(-)}
&=&
\frac{\hbar^2}{2}
\frac{(A_{k,l}^{(2)})^2}{\omega_k \omega_{l}}
(n_k +n_{l} + 2 n_k n_{l}),
\\
A_{k,l}^{(2)}
&=& -3 G_{S,k,l},
\\
n_k &=& 1/(e^{\beta \hbar \omega_k}-1). 
\end{eqnarray}

\subsection{Maradudin-Fein formula}

There exists another well known formula to describe 
the VER rate, the Maradudin-Fein (MF) formula \cite{MF62,Leitner01},
\begin{eqnarray}
W &=& W_{\rm decay} + W_{\rm coll},
\label{eq:MF}
\\
W_{\rm decay} 
&=&
\frac{\hbar}{2 m_S \omega_S} 
\sum_{k,l} \frac{(A_{k,l}^{(2)})^2}{\omega_k \omega_{l}}
(1+n_k +n_{l}) \frac{\gamma}{\gamma^2+(\omega_S -\omega_k-\omega_{l})^2},
\\
W_{\rm coll}
&=&
\frac{\hbar}{m_S \omega_S} 
\sum_{k,l} \frac{(A_{k,l}^{(2)})^2}{\omega_k \omega_{l}}
(n_k -n_{l}) \frac{\gamma}{\gamma^2+(\omega_S +\omega_k-\omega_{l})^2}
\end{eqnarray}
with a width parameter $\gamma$.
Note that Eq.~(\ref{eq:rate}) and Eq.~(\ref{eq:MF}) are equivalent 
in the limit of $\gamma \rightarrow 0$
as shown by Kenkre, Tokmakoff, and Fayer \cite{KTF94}.
However, they disagree with a finite width parameter such as $\gamma \sim 100$ cm$^{-1}$.
In this chapter, we use the MF formula and consider 
its classical limit ($\hbar \rightarrow 0$) defined as 
\begin{eqnarray}
W^{\rm cl}_{\rm decay} 
&=&
\frac{1}{2 m_S \beta \omega_S} 
\sum_{k,l} \frac{(A_{k,l}^{(2)})^2}{\omega_k \omega_{l}}
\left( \frac{1}{\omega_k} +\frac{1}{\omega_l} \right)
\frac{\gamma}{\gamma^2+(\omega_S -\omega_k-\omega_{l})^2},
\\
W^{\rm cl}_{\rm coll}
&=&
\frac{1}{m_S \beta \omega_S} 
\sum_{k,l} \frac{(A_{k,l}^{(2)})^2}{\omega_k \omega_{l}}
\left( \frac{1}{\omega_k} -\frac{1}{\omega_l} \right)
\frac{\gamma}{\gamma^2+(\omega_S +\omega_k-\omega_{l})^2}.
\end{eqnarray}
We note some properties of the formula:
$W_{\rm decay} \geq W^{\rm cl}_{\rm decay}$ and 
$W_{\rm coll} \leq W^{\rm cl}_{\rm coll}$ 
which is derived from 
\begin{eqnarray}
\frac{1/(e^x-1)+1/(e^y-1)}{1/x+1/y} &\geq& 1,
\\
\frac{1/(e^x-1)-1/(e^y-1)}{1/x-1/y} &\leq& 1
\end{eqnarray}
for $x,y >0$.
We can define an effective QCF as 
\begin{equation}
Q_{\rm eff}=\frac{W}{W^{\rm cl}}
=\frac{W_{\rm decay} + W_{\rm coll}}{W^{\rm cl}_{\rm decay} + W^{\rm cl}_{\rm coll}}.
\label{eq:eff}
\end{equation}
This should be compared with the normalized QCFs 
[$\tilde{Q}=Q(\omega_S)/(\beta \hbar \omega_S)$] found in the literature.

\subsection{Third order coupling elements}

To apply the Maradudin-Fein theory to the case of CD bond relaxation in cytochrome c, we must numerically compute the third order coupling elements.  As the protein has more than $10^3$ modes, there are more than $10^6$ third order coupling elements representing the coupling of the ``system'' CD bond to the ``bath'' modes of the surrounding protein and solvent.  

We have employed the finite difference approximation
\begin{equation}
A^{(2)}_{mn}=-\frac{1}{2} \frac{\partial^3 V}{\partial q_S \partial q_m \partial q_n}
\simeq -\frac{1}{2} \sum_{ij} U_{im}U_{jn} 
\frac{K_{ij}(\Delta q_S)-K_{ij}(-\Delta q_S)}{2\Delta q_S}
\end{equation}
where $U_{ik}$ is an orthogonal matrix that diagonalizes 
the (mass-weighted) hessian matrix at the mechanically stable structure $K_{ij}$,
and $K_{ij}(\pm \Delta q_{S})$ is a hessian matrix 
calculated at a shifted structure along the direction of 
a selected mode with a shift $\pm \Delta q_{S}$.

Note that, in the large number of coupling elements, 
most are small in magnitude.  
Of those that are larger, 
most fail to meet the resonance condition and do not contribute significantly 
to the perturbative estimate of the VER rate. 
See \cite{FBS04} for the details.

\subsection{Width parameter}

\begin{figure}[htbp]
\hfill
\begin{center}
\begin{minipage}{.42\linewidth}
\includegraphics[scale=1.1]{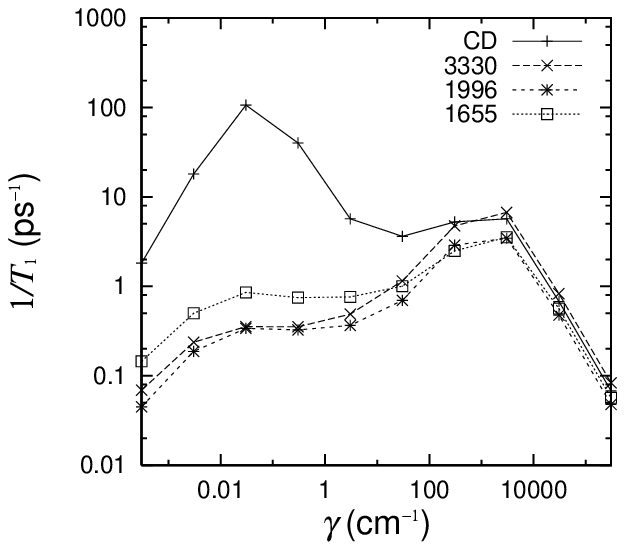}
\end{minipage}
\hspace{1cm}
\begin{minipage}{.42\linewidth}
\includegraphics[scale=1.1]{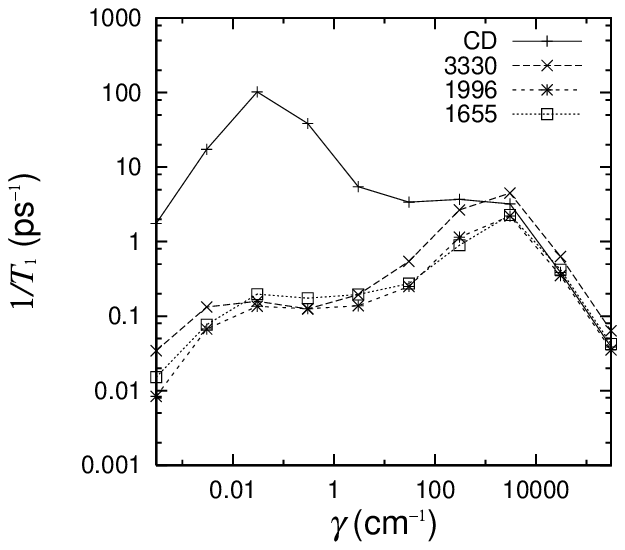}
\end{minipage}
\end{center}
\caption{
VER rates 
for the CD mode 
($\omega_{CD}=2129.1$ cm$^{-1}$) 
and the other lower frequency modes 
($\omega_{3330}=1330.9$ cm$^{-1}$, $\omega_{1996}=829.9$ cm$^{-1}$, $\omega_{1655}=685.5$ cm$^{-1}$)
as a function of $\gamma$ at 300K (left) and at 15K (right).
}
\label{fig:VER}
\end{figure}

We show the width parameter $\gamma$ 
dependence of the VER rate in Fig.~\ref{fig:VER}.\footnote{
Note two limiting cases of $\gamma$ dependence: 
$1/T_1 \propto \gamma$ when $\gamma$ is very small,
and $1/T_1 \propto 1/\gamma$ when $\gamma$ is very large.
This is easily recognized from the Lorentzian form.
}
We consider other lower 
frequency modes 
($\omega_{3330}=1330.9$ cm$^{-1}$, $\omega_{1996}=829.9$ cm$^{-1}$, $\omega_{1655}=685.5$ cm$^{-1}$)
as well as 
the CD mode ($\omega_{CD}=2129.1$ cm$^{-1}$) 
for comparison. From the former analysis of the 
frequency autocorrelation function Eq.~(\ref{eq:fre-auto}), 
we might be able to take $\gamma \simeq \Delta \omega \sim 3$ cm$^{-1}$ for the CD mode,
and we have $T_1 \simeq 0.2$ ps, which agrees with the previous result with QCFs: 
$T_1^{\rm QCF} = 0.3 \sim 0.4$ ps.

We also see that the lower frequency modes have longer VER time, a few ps,
which agrees with the calculations by Leitner's group employing the MF formula \cite{Leitner04}.
The main contribution to the VER rate at $\gamma =3$ cm$^{-1}$ comes 
from 1655th mode (685.5 cm$^{-1}$), a heme torsion, and the 3823rd (1443.5 cm$^{-1}$) mode, an angle bend in Met80 ($\sim 20\%$).
Interestingly, we can conceive a peak around $\gamma =0.03$ cm$^{-1}$.
Given this width parameter, the contribution from the two modes is more than 90\%.
In any case, we can say that  1655th and 3823rd modes are resonant with the CD mode
because they satisfy the resonant condition 
($|\omega_{1655}+\omega_{3823}-\omega_{CD}| \simeq 0.03$ cm$^{-1}$)
and the coupling elements between 
them is relatively large ($|A^{(2)}_{1655,3823}| \simeq 5.1$ kcal/mol/\AA$^3$).  

However, this close resonance does not necessarily lead to the conclusion that it forms the dominant channel for VER of the CD strech.  There is a competing near-resonance involving the 3330th mode (1330.9 cm$^{-1}$), an angle bending mode of Met80, and the 1996th mode (829.9 cm$^{-1}$), a stretch-bend mode in Met80.   While the resonance is not close  (it is within 31.7 cm$^{-1}$), the coupling element is quite large  ($|A^{(2)}_{3330,1996}| \simeq 22.3$ kcal/mol/\AA$^3$).  With a larger value of $\gamma=30 {\rm cm}^{-1}$, this combination of bath modes becomes the dominant channel for VER of the CD stretch.  Clearly, the uncertainty in our force field, used to compute the vibrational frequencies, and the value of $\gamma$, which is rather poorly defined, prevents us from concluding that one or another of these two channels will dominate VER of the CD stretch at room temperature.

\begin{figure}[htbp]
\hfill
\begin{center}
\begin{minipage}{.42\linewidth}
\includegraphics[scale=1.1]{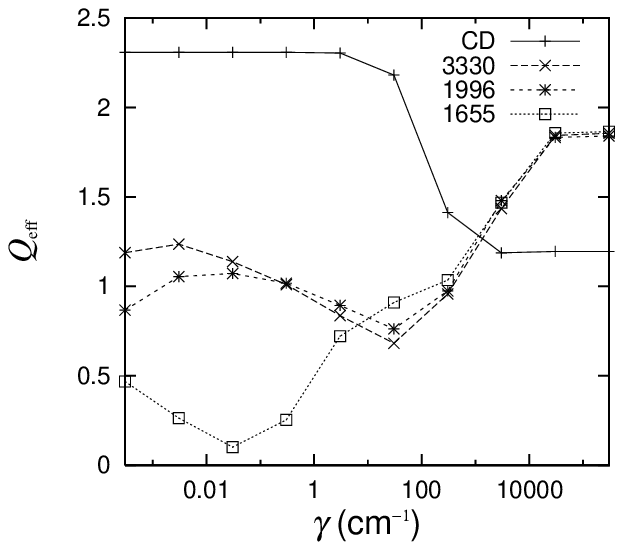}
\end{minipage}
\hspace{1cm}
\begin{minipage}{.42\linewidth}
\includegraphics[scale=1.1]{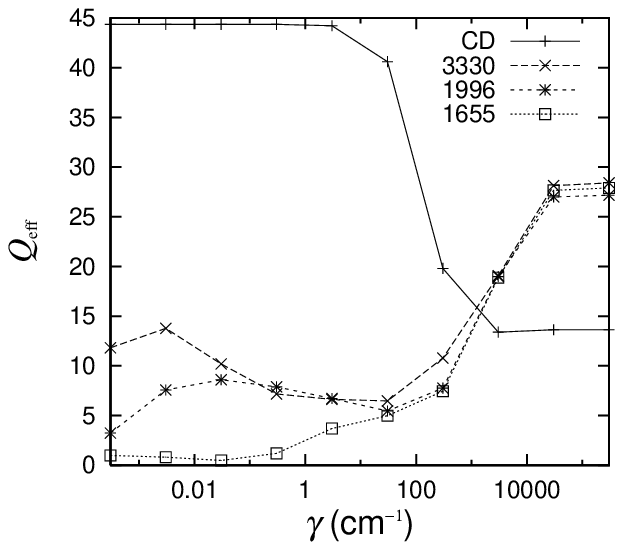}
\end{minipage}
\end{center}
\caption{
Effective QCF 
for the CD mode 
($\omega_{CD}=2129.1$ cm$^{-1}$) 
and 
the other lower frequency modes 
($\omega_{3330}=1330.9$ cm$^{-1}$, $\omega_{1996}=829.9$ cm$^{-1}$, $\omega_{1655}=685.5$ cm$^{-1}$)
as a function of $\gamma$ at 300K (left).
and at 15K (right).
}
\label{fig:ratio}
\end{figure}

In the left of Fig.~\ref{fig:ratio}, we show the effective QCF calculated 
from Eq.~(\ref{eq:eff}) at 300K,
which is $Q_{\rm eff} \simeq 2.3$ for the CD mode with $\gamma=3$ cm$^{-1}$.
This value better agrees with the (normalized) harmonic-harmonic QCF [Eq.~(\ref{eq:QCFHH})],
compared to the harmonic-harmonic-Schofield QCF [Eq.~(\ref{eq:QCFHHS})].
The $Q_{\rm eff}$ for the other modes are more or less unity, which indicates 
that these modes behave classically at 300K.\footnote{
We notice an interesting behavior for 1655th mode, i.e. $Q_{\rm eff}$ becomes 
very much smaller than unity at $\gamma \simeq 0.03$ cm$^{-1}$.
In this case, we observe that $W_{\rm coll} \gg W_{\rm decay}$ because of the 
resonance: $\omega_{1655}+\omega_{3823}-\omega_{CD} \simeq 0$ (actually 0.03 cm$^{-1}$).
In such a case, $Q_{\rm eff}$ becomes less than unity because 
$W_{\rm coll} \leq W^{\rm cl}_{\rm coll}$.
}
In contrast, as is shown in the right of Fig.~\ref{fig:ratio}, 
$Q_{\rm eff}$ at 15K is very large ($Q_{\rm eff} \simeq 40$),
which implies that the classical VER rate becomes small because 
it is proportional to the temperature (see also Fig.~\ref{fig:temp}).
A similar trend is found in the right of Fig.~\ref{fig:QCF}, 
where the harmonic-harmonic QCF ($\tilde{Q} \simeq 40$) is comparable to $Q_{\rm eff}$.
On the other hand,
the harmonic-harmonic-Schofield QCF gives an exponentially large value of $\tilde{Q}$,
showing strong deviations from $Q_{\rm eff}$.
We should bear in mind that different QCFs lead to significantly different 
conclusions at low temperatures.
  
\subsection{Temperature dependence}

\begin{figure}[htbp]
\hfill
\begin{center}
\begin{minipage}{.42\linewidth}
\includegraphics[scale=1.1]{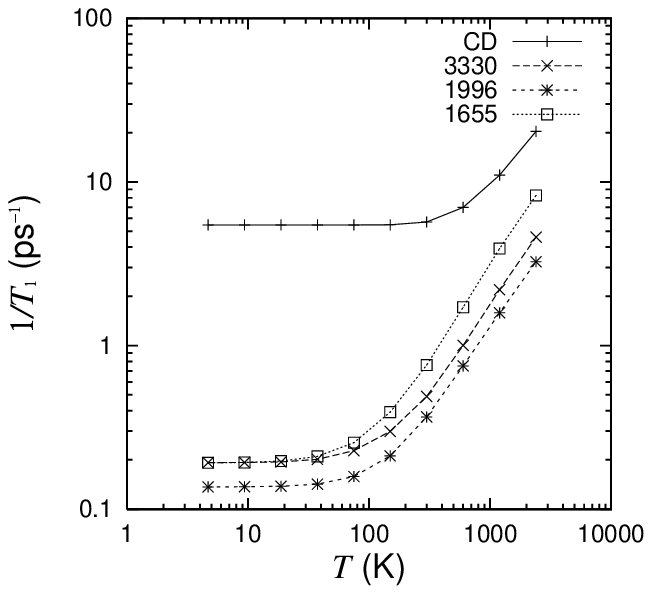}
\end{minipage}
\hspace{1cm}
\begin{minipage}{.42\linewidth}
\includegraphics[scale=1.1]{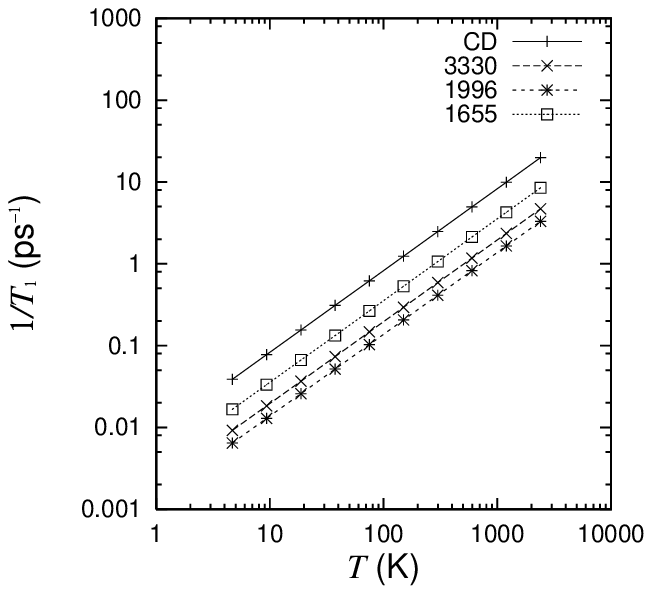}
\end{minipage}
\end{center}
\caption{
Quantum (left) and classical (right) 
VER rates for the CD mode 
($\omega_{CD}=2129.1$ cm$^{-1}$) 
and 
the other lower frequency modes 
($\omega_{3330}=1330.9$ cm$^{-1}$, $\omega_{1996}=829.9$ cm$^{-1}$, $\omega_{1655}=685.5$ cm$^{-1}$)
as a function of temperature with $\gamma=3$ cm$^{-1}$.
}
\label{fig:temp}
\end{figure}

In Fig.~\ref{fig:temp}, we show the temperature dependence of 
the quantum and classical VER rate calculated by the MF formula and 
the classical limit of the MF formula.
At high tempratures ($\sim 1000$K), 
the quantum VER rate agrees with the classical one,
but they deviate at low temperatures.
The former becomes constant due to the remaining quantum fluctuation 
(zero point energy)
whereas the latter decreases as $\propto$ (temperature).
The ``cross over temperature'' where the VER behaves classically 
is smaller for the lower frequency modes compared to that of 
the CD mode as expected.

\section{Discussion}

\subsection{Comparison with experiment}

Here we compare our results with the experiment by Romeberg's group \cite{CJR01}.
They measured the shifts and widths of the spectra 
for different forms of cyt c;
the widths of the spectra (FWHM) 
were found to be $\Delta \omega_{\rm FWHM} \simeq 6.0 \sim 13.0$ cm$^{-1}$.
From the discussions of Sec.~\ref{sec:freq-fluc},
we can theoretically neglect inhomogeneous effects, 
and estimate the VER rate simply as 
\begin{equation}
T_1 \sim 5.3/\Delta \omega_{\rm FWHM}  \,({\rm ps})
\label{eq:estimate}
\end{equation}
which corresponds to $T_1 \simeq 0.4 \sim 0.9$ ps.
This estimate is similar to the QCF prediction 
using Eq.~(\ref{eq:qcf}) (0.3 $\sim$ 0.4 ps)
and the reduced model approach using Eq.~(\ref{eq:rate}) or (\ref{eq:MF}) (0.2 $\sim$ 0.3 ps).
We note that this value should be compared to the VER time 
of the C-H stretch in N-methylacetamide-D [CH$_3$(CO)ND(CH$_3$)] \cite{HLH98},
which is also sub ps. 
Further experimental studies, e.g. on temperature dependence of absorption spectra or 
on time-resolved spectroscopy, will clarify which methodology is more applicable.

Romesberg's group studied
Met80-3D (methionine with three deuteriums) 
while we have examined Met80-1D (methionine with one deuterium).
In the case of Met80-3D, there are three peaks in the transparent 
region.  It should be possible to consider the VER of each of the three modes, to make a more direct comparison of the predictions of our theoretical models with the results of their experimental studies.


\subsection{Validity of Fermi's golden rule}

We next discuss the validity of our approaches.
Since our starting point is the perturbative Fermi's golden rule, 
our two approaches should have a limited range of validity.
Naively speaking, the force applied on the CD mode should be small 
enough, but how small it should be?

We follow Kubo's derivation of a quantum master equation using 
the projection operator technique \cite{Kubo}.
He derived an equation for the evolution
of the system density operator $\sigma(t)$
\begin{eqnarray}
\frac{\partial}{\partial t} \sigma(t)\
&=&
-\frac{1}{\hbar^2}
\int_{-\infty}^t d \tau 
[q(t)q(\tau) \sigma(\tau) \Phi(t-\tau)
-q(t) \sigma(\tau) q(\tau) \Phi(-t+\tau)
\nonumber
\\
&&
+\sigma(\tau) q(\tau) q(t) \Phi(-t+\tau)
-q(\tau) \sigma(\tau) q(t) \Phi(t-\tau)].
\label{eq:master}
\end{eqnarray}
The interaction Hamiltonian is assumed to be 
${\cal H}_{\rm int}=- q {\cal F}$, as in our case, i.e. $q$ is the system 
coordinate and ${\cal F}$ mainly contains the bath coordinates.
We have defined the force autocorrelation function  
$\Phi(t)={\rm Tr}_B \{ \rho_B {\cal F}(t) {\cal F}(0) \} 
= \langle {\cal F}(t) {\cal F}(0) \rangle$.
Note that 
Eq.~(\ref{eq:master}) is just a von Neumann equation using the projection operator technique,
and it is not a master equation yet.

If $\Phi(t)$ decays fast, we can replace $\sigma(\tau)$ in the integral 
with $\sigma(t)$, and the dynamics becomes an approximate 
Markovian dynamics. If this approximation is valid, 
Fermi's golden rule describes the 
relaxation dynamics of $\sigma(t)$ \cite{Kubo}.
The validity of the golden rule relies on the validity of 
the Markov approximation.

From Eq.~(\ref{eq:master}), 
the relaxation rate of $\sigma(t)$ can be estimated as 
\begin{equation}
1/\tau_r \sim (\langle q^2 \rangle {\cal F}^2/\hbar^2) \tau_c
\end{equation}
where we have assumed 
\begin{equation}
\Phi(t) \simeq {\cal F}^2 e^{-|t|/\tau_c}.
\end{equation}
The Markov approximation [$\sigma(\tau) \simeq \sigma(t)$] holds for 
\begin{equation}
\tau_r \gg \tau_c.
\end{equation}
We have a criterion for the validity of the Markov approximation 
\begin{equation}
\epsilon \equiv \langle q^2 \rangle {\cal F}^2  \tau_c^2 /\hbar^2 \ll 1.
\label{eq:Kubo}
\end{equation}
In our case as well as the case of HOD in D$_2$O, 
the ratio is ``just'' small (see Table \ref{tab:ratio}). 
Applying Fermi's golden rule to these situations 
should be regarded as a reasonable estimate of the VER rate.
As alternative approaches that avoid this underlying Markov
approximation, one can employ more ``advanced'' methods 
as mentioned in Sec.~\ref{sec:summary}.

\begin{table}[htbp]
\caption{
The parameters in Eq.~(\ref{eq:Kubo}) for various molecules in AKMA units
(unit length = 1 \AA, unit time = 0.04888 ps, unit energy = 1 kcal/mol). 
The data for HOD in D$_2$O, 
CN$^-$ in water, and CO in Mb 
are taken from \cite{RH96}, \cite{RH98}, and \cite{SS99}, respectively. 
}
\hfill
\begin{center}
\begin{tabular}{c|c|c|c|c}
\hline \hline
  & $\langle q^2 \rangle$ & ${\cal F}^2$ & $\tau_c$ & $\epsilon$ \\ \hline \hline
CD in cyt c & 0.01 & 5.0 & 1.0 & 0.5 \\ \hline
HOD in D$_2$O  & 0.01 & 5.0 & 1.0 & 0.5 \\ \hline
CN$^{-}$ in water & 0.002 & 1.0 & 0.5 & 0.01 \\ \hline
CO in Mb & 0.002 & 1.0 & 1.0 & 0.02 \\ \hline
\hline
\end{tabular}
\end{center}
\label{tab:ratio}
\end{table}

\subsection{Higher order coupling terms}

Up to now, we have only included the third order coupling terms
to describe the VER of the CD mode. However, 
we must be concerned with the relative contribution of 
higher order mechanism, e.g. the contribution due to the fourth order coupling terms in Eq.~(\ref{eq:4th}).
This is a very difficult question. 
As there are many terms ($\sim 10^9$) included,
we cannot directly calculate all of them for cyt c.
We have found that it is not sufficient to include only the 
third order coupling terms to reproduce the fluctuation of the force on 
the CD bond. However, this does not necessarily mean that the VER rate 
calculated from the third order coupling terms is inadequate.

The main contribution from the fourth order 
coupling terms to the VER rate and the force fluctuation are  written as 
\begin{equation}
\Delta \left( \frac{1}{T_1} \right)
\sim
\sum_{k,l,m} \frac{|H_{S,k,l,m}|^2}{\omega_k \omega_l \omega_m}
\delta(\omega_S - \omega_k -\omega_l -\omega_m),
\end{equation}
and 
\begin{equation}
\Delta \langle \delta {\cal F}^2 \rangle
\sim
\sum_{k,l,m} \frac{|H_{S,k,l,m}|^2}{\omega_k \omega_l \omega_m},
\end{equation}
respectively.
Even if $\Delta \langle \delta {\cal F}^2 \rangle$ becomes large,
$\Delta \left( 1/T_1 \right)$ is not necessarily large 
because of the resonance condition ($\omega_S - \omega_k -\omega_l -\omega_m \simeq 0$).
It is a future task how to evaluate the effects due to higher order coupling terms.
It might be interesting to compare the classical limit using the reduced model 
approach and the Landau-Teller-Zwanzig approach 
because there is no ambiguity about how to choose QCFs \cite{Okazaki}.

\section{Summary}
\label{sec:summary}

In this chapter, we have examined VER in a protein from the QCF approach
and the reduced model approach, and compared the results. 
For the CD mode in cyt c (in vacuum) at room temperature, 
both approaches yield the same result for the VER rate, 
which is also very similar to an estimate based on an experiment by Romesberg's group.
Our work demonstrates both the feasibility and accuracy of a number of theoretical approaches 
to estimate VER rates of selected modes in proteins.

The QCF approach is appealing in that the calculation of the force autocorrelation function is straightforward and feasible, even for systems of thousands of degrees of freedom.  Moreover, the classical force autocorrelation function includes all orders of non-linearity in the interaction between the system oscillator and the surrounding bath.  A weakness of the QCF approach is that 
we don't know which QCF to choose {\it a priori}.   We must assume a mechanism for VER before computing the rate.   Moreover, the temperature dependence of the rate of VER is sensitive to the mechanism, whether it involves few phonons or many phonons. The choice of the form of the quantum correction factor can make a significant difference in the predicted rate of VER at lower temperatures.

On the other hand,
the reduced model approach is appealing in that the quantum dynamics of the reduced system is accurately treated.  Using the reduced model approach, there are two ways to estimate the quantum mechanical force autocorrelation function: (1) numerical calculation of the quantum dynamics for a model Hamiltonian of a few degrees of freedom, including all orders of non-linearity in the potential, and (2) analytical solution for the quantum dynamics using perturbation theory that includes many bath modes but only the lowest order non-linear coupling between system and bath modes.   We have employed the latter approach through the use of the Maradudin-Fein formula.   A weakness of our reduced model approach is 
that the method neglects the higher order coupling 
elements beyond third order, which cannot be justified {\it a priori} \cite{MSO01,GCGB03}.

We have pursued a comparative study in which we seek {\it consensus} in the estimates of $1/T_1$ that result of the QCF approach and the perturbation theory.
A rather remarkable result of our study is that while the absolute value of the quantum corrections to the classical VER theory are large (on the order of a factor of 40), the results of the QCF and the perturbation theory approaches are in close agreement.  This is all the more remarkable given the fact that the results of the perturbation theory require a calculation of the third-order coupling constants and the estimation of the ``lifetime'' parameter $\gamma$.  As we have shown, the dominant channel for VER, derived from the perturbation theory, depends upon the choice of $\gamma$.  For smaller $\gamma = 3 {\rm cm}^{-1}$, the dominant mechanism is a close resonance (within 0.1 cm$^{-1}$), a combination of a heme torsion and Met80 angle bending mode, with a weak coupling.  For larger $\gamma = 30 {\rm cm}^{-1}$, the dominant mechanism appears to be a less perfect resonance (within 31.7 cm$^{-1}$), a combination of a different angle bending mode and a bend-stretch mode in Met 80, with a strong coupling.  Such detailed knowledge of $\gamma$ is essential to predict a mechanism for VER. 

Our study raises two important questions: 
(1) What is the optimal set of coordinates for modeling and interpreting VER in proteins?  
In the QCF approach, 
we treated the relaxing bond (CD bond) as a {\it local mode} that is coupled 
to vibrational modes of the bath.
In the reduced model approach, on the other hand, 
we treated all the vibrational modes including the relaxing mode (CD mode) 
as {\it normal modes} that are coupled each other with the 
third order nonlinear coupling terms.
Our numerical results showed that the two approaches give similar results for 
the VER rate of the CD bond or mode,
but it remains to be seen that this is a kind of coincidence or 
there is a theoretical ground of their equivalence (if the QCF is appropriately chosen).
(2) What is the physical origin of the width parameter $\gamma$ and how to calculate it?
In this work,
we suggested to use the relation $\gamma \simeq \Delta \omega$ where 
$\Delta \omega$ represents the fluctuation of the CD mode (or bond) frequency.
We think this is reasonable but there is no theoretical explanation of this.
If the VER rate does not significantly depend on $\gamma$, this is not 
a serious problem, but this is not always the case.
Thus we need an ``ab initio'' way to derive the width parameter $\gamma$. 
One appealing way is to regard $\gamma$ as a hopping rate between 
potential basins (inherent structures) \cite{MKS91,CV95}.
The other is the more rigorous quantum mechanical treatment of the 
tier structure of energy levels in the protein \cite{SM93}.


The results of our study are derived through the use of an approximate empirical energy function (force field) which has not been ``tuned'' to provide accurate frequencies of vibration for all protein modes.  Our predicted rates of VER depend sensitively on the closeness of the resonance between the system and bath modes.  Clearly, we must resort to the re-parameterization of the empirical potential to fit with experimental data or higher levels of theory (ab initio quantum chemistry calculation) in an effort to refine our estimates of the frequencies of vibration and the details of non-linear coupling between vibrational modes of the protein.  This is a challenge for both experimental and theoretical studies.

Recent advances in experiment and theory make the present time an exciting one for the detailed study of protein dynamics. A variety of methods have been applied to examine VER in molecules, including nonequilibrium MD methods \cite{NS03}, time-dependent self-consistend field methods \cite{GCGB03,RGER95}, mixed quantum-classical methods \cite{TSO01}, 
and semiclassical methods \cite{SG03a,Rossky}.
In addition, it is now possible to compute spectroscopic observables such 
as absorption spectra or 2D-IR signals \cite{MA04,KT01} as probes of protein structure and dynamics.  Extensions of these studies will provide us with an increasingly detailed picture of the dynamics of proteins, and its relation to structure and function.

\section{Acknowledgements}

We thank Prof. S. Takada, Prof. T. Komatsuzaki,
Prof. Y. Mizutani, Prof. K. Tominaga, Prof. S. Okazaki, 
Prof. F. Romesberg,
Prof. J.L. Skinner, Prof. D.M. Leitner,
Prof. I. Ohmine, Prof. S. Saito, Prof. R. Akiyama,
Prof. K. Takatsuka, Dr. H. Ushiyama,
Dr. T. Miyadera, Dr. S. Fuchigami,
Dr. T. Yamashita, Dr. Y. Sugita,
Dr. M. Ceremeens, Dr. J. Zimmerman, and Dr. P.H. Nguyen 
for helpful comments and discussions,
and Prof. S. Mukamel and Prof. Y. Tanimura for informing the 
references related to nonlinear spectroscopy.  We thank the National Science Foundation (CHE-0316551) for its generous support of our research.

\end{document}